\documentclass[groupedaddress,jap,aip,preprint,graphicx]{revtex4-1} % for review purposes
\usepackage{graphicx}
\usepackage{epstopdf}
\usepackage{times}
\usepackage[small, compact]{titlesec}

\begin{document}
\title{Occurrence of Mixed Phase in $\bf{Bi_{0.5}Sr_{0.5}Mn_{0.9}Cr_{0.1}O_3}$ bulk sample: Electron Paramagnetic Resonance and Magnetization Studies} 
\author{Bhagyashree K. S.} 
\author{Lora Rita Goveas$^*$}
\author{S. V. Bhat}
\affiliation{Department of Physics, Indian Institute of Science, Bangalore-560012, India}
\affiliation{$^*$St. Josephs College (Autonomous), Bangalore 560027, India}

\begin{abstract}
We study the effects of 10\% Cr substitution in Mn sites of $Bi_{0.5}Sr_{0.5}MnO_3$ on the antiferromagnetic (AFM) ($T_N \sim$ 110 K) transition using structural, magnetic and electron paramagnetic resonance (EPR) techniques. Field cooled (FC) and zero field cooled (ZFC) magnetization measurements done from 400 K down to 4 K show that the compound is in the paramagnetic (PM) phase till 50 K where it undergoes a transition to a short range ferromagnetic phase (FM). Electron paramagnetic resonance measurements performed in the temperature range 300 K till 80 K conform with the magnetization measurements as symmetric signals are observed owing to the paramagnetic phase. Below 80 K, signals become asymmetric. Electron paramagnetic resonance intensity peaks at $\sim$ 110 K, the decreasing intensity below this temperature confirming the presence of antiferromagnetism. We conclude that below 50 K the magnetization and EPR results are consistent with a cluster glass phase of BSMCO, where ferromagnetic clusters coexist with an antiferromagnetic background. 
\end{abstract}

%\begin{keyword}
% EPR \sep Manganites \sep CO \sep AFM \sep FM
%\end{keyword}

\maketitle
\section{INTRODUCTION}
Manganites of the form $T_{1-x}D_xMnO_3$ (where ‘T’ is a trivalent rare-earth ion or ‘Bi’ ion and ‘D’ is a divalent alkaline earth ion) display competition between ferromagnetic double exchange (Zener exchange) interaction (between $Mn^{3+}$ and $Mn^{4+}$ ions leading to ferromagnetic metallic (FMM) state) and antiferromagnetic (AFM) super-exchange interaction (between $Mn^{3+}$ or $Mn^{4+}$ ions leading to charge ordered insulating (COI) AFM state)~\cite{tokura,CNRrao,Rama}. This competition leads to the formation of domains with contrasting properties, i.e. the system stabilizes by breaking into regions with high electron density (this region becomes AFM) and low electronic density (this region becomes FM). This phenomena is called phase separation~\cite{Dagotto}. Coulomb repulsion between electrons suggests that these regions are of a few nanometers in size. The end members of the phase diagram in these compounds ($TMnO_3$ and $DMnO_3$) are stabilized in AFM state; when the end members are doped with divalent ions in case of $TMnO_3$ and trivalent ions in $DMnO_3$, FM state starts to build up. This leads to the formation of mixed phase with AFM phase and FM phase existing together. Even in the half doped $T_{0.5}D_{0.5}MnO_3$, COI-AFM state is stabilized due to the presence of equal number of $Mn^{3+}$ and $Mn^{4+}$ ions. Interestingly, it has been found that a few methods like external pressure, magnetic field, reducing the size of the bulk particle to nano and substitution of the ‘$Cr^{3+}$’ ions in sites of ‘$Mn^{3+}$’ ions have resulted in suppression of the CO-AFM phase and enhancement of FMM phase~\cite{Anu,Tripathi,Sharma,Xiong}. ‘$Cr^{3+}$’  breaks the long range CO state due to its unoccupied $e_g$ orbital, suppressing the AFM phase and enhancing the double exchange interaction resulting in the stabilization of FMM phase. 

A few investigations on Cr doped rare-earth manganites have been reported previously~\cite{Sharma,Dho,Troy1,Troy}. Studies done on $Nd_{0.5}Ca_{0.5}Mn_{1-y}Cr_yO_3$ (y=0.03, 0.05, 0.1)~\cite{Sharma} show that with increasing Cr substitution CO AFM phase diminishes and FMM phase emerges. Studies done on $La_{0.46}Sr_{0.54}Mn_{1-y}Cr_yO_3$ ($0\leq y \leq 0.08$)~\cite{Dho} show that at y = 0 the sample undergoes PM to FM and FM to AFM transitions at 272 K and 190 K respectively. With increasing Cr substitution the FM phase becomes dominant over AFM phase. At y = 0.08 AFM phase disappears completely. Studies done on $Nd_{0.6}Ca_{0.4}Mn_{1-y}Cr_yO_3$ system~\cite{Troy1} show that, the parent compound which is CO AFM
changes to a mixed magnetic state with AFM and FM domains coexisting in the region $0.015 \leq y \leq 0.04$ of Cr substitution. In the region $0.04 < y < 0.4$  the sample is in the ferrimagnetic state. Neutron diffraction and magnetization studies done on $Nd_{0.6}Ca_{0.4}Mn_{0.5}Cr_{0.5}O_3$~\cite{Troy} show that manganese and chromium ions order in the lattice in a similar way to $Mn^{3+}$ and $Mn^{4+}$ ordering in the CO state of $Nd_{0.6}Ca_{0.4}MnO_3$. Below 100 K the magnetic moment increases with decreasing temperature and does not saturate, indicating the presence of FM component. Takuro et.al.~\cite{Takuro} studied the effects of Cr substitution as an impurity in $Pr_{1-x}Ca_xMn_{0.97}Cr_{0.03}O_3$, $La_{1-x}Ca_xMn_{0.97}Cr_{0.03}O_3$ and $Nd_{1-x}Sr_xMn_{0.97}Cr_{0.03}O_3$. In here just 3\% of Cr substitution has induced a short range CO state, but the FM and layered AFM phase remain as long range ordered.

Bismuth based manganites of the formula $Bi_{1-x}D_xMnO_3$ (where D=Ca,Sr) are different from doped rare-earth manganites due to the presence of highly polarizable $6s^2$ lone pair of electrons present on the ‘Bi’ ion. It significantly restricts the mobility of $e_g$ electrons and favors charge ordering. Very few investigations have been reported on ‘Cr’ substituted Bi manganites~\cite{Xiong,Yamada}. $Bi_{0.5}Sr_{0.5}MnO_3$ shows $T_{CO} \sim$ 525 K and $T_N \sim$ 110 K~\cite{Herv,Kirste}. CO in this composition is very robust and this is also followed by the AFM phase. Rao et.al~\cite{Rao} analyzed $Bi_{0.5}Sr_{0.5}MnO_3$ nanoparticles and found that both CO and AFM phases are unaffected unlike the other doped rare-earth manganites~\cite{Anu,Tripathi}. Aim of this work is to study if 10\% of Cr substitution is sufficient to modify the existing properties and give rise to new properties in $Bi_{0.5}Sr_{0.5}MnO_3$ system. We discovered that Cr substitution has led to the creation of short range FM phase embedded in the matrix of AFM phase in the back ground.

\section{PREPARATION AND CHARACTERIZATION}

\begin{figure}[hbt!]
\includegraphics[width=\linewidth]{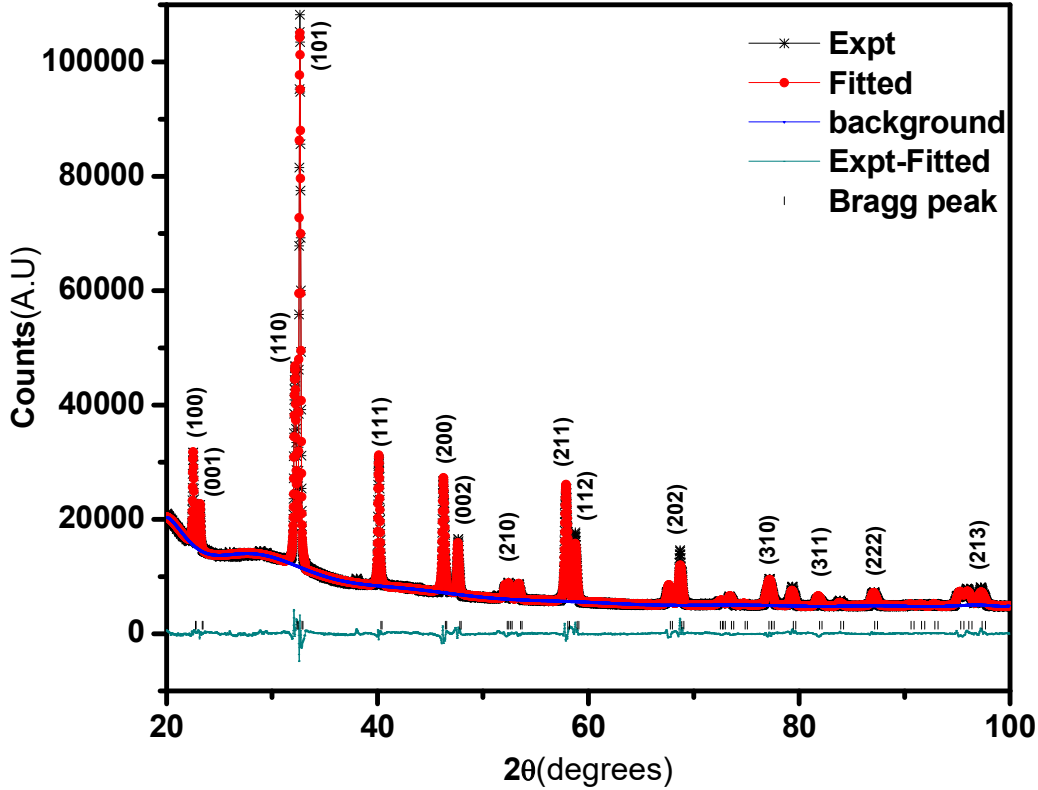}
\caption{Rietveld refined XRD of BSMCO sample; Here black cross represents experimental data, red circles represents the fit, blue triangle represents background, green line represents the difference between experimental and fitted data, and the vertical bars are the Bragg reflection positions.}
\label{XRD}
\end{figure}

\begin{figure}[hbt!]
\includegraphics[width=\linewidth]{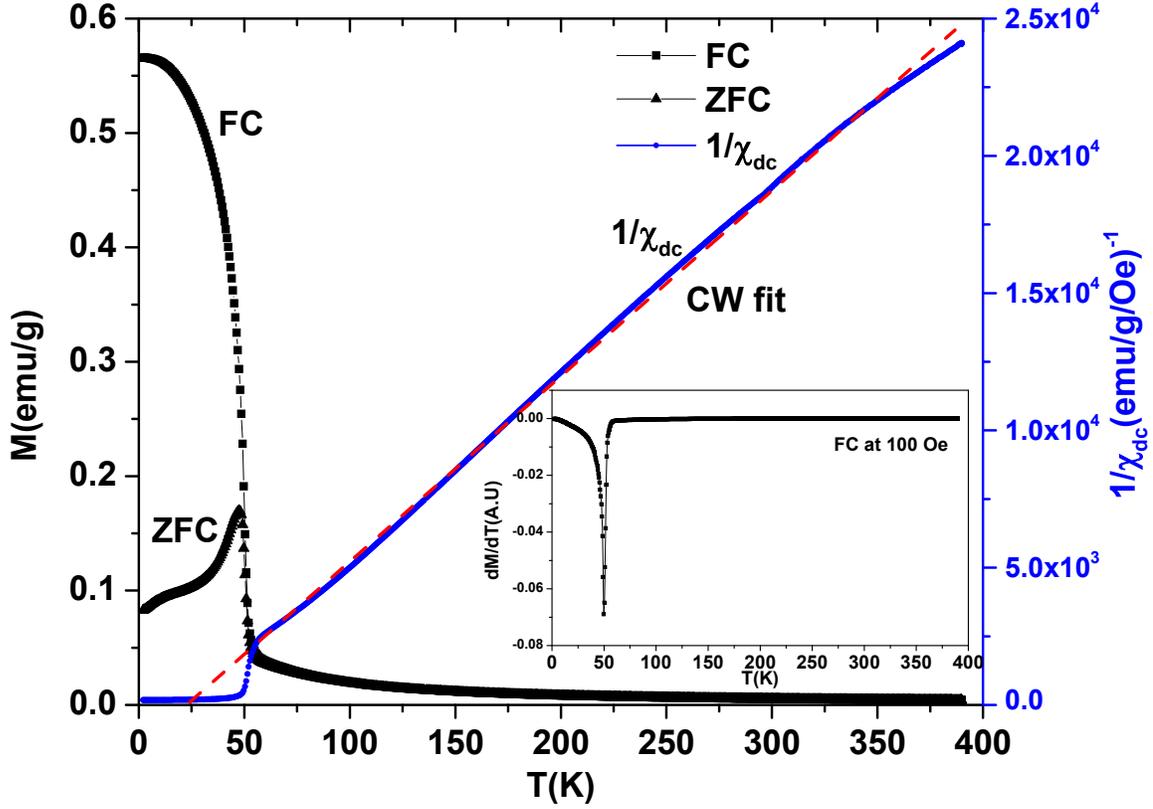}
\caption{FC, ZFC and $1/\chi$ behavior of BSMCO sample in the presence of 100 Oe field; The dashed line is the Curie-Weiss fit to the experimental data which look like a continuous line because of the large number of data points. Inset shows the dM/dT plot.}
\label{FZ}
\end{figure}

\begin{figure}[hbt!]
\includegraphics[width=\linewidth]{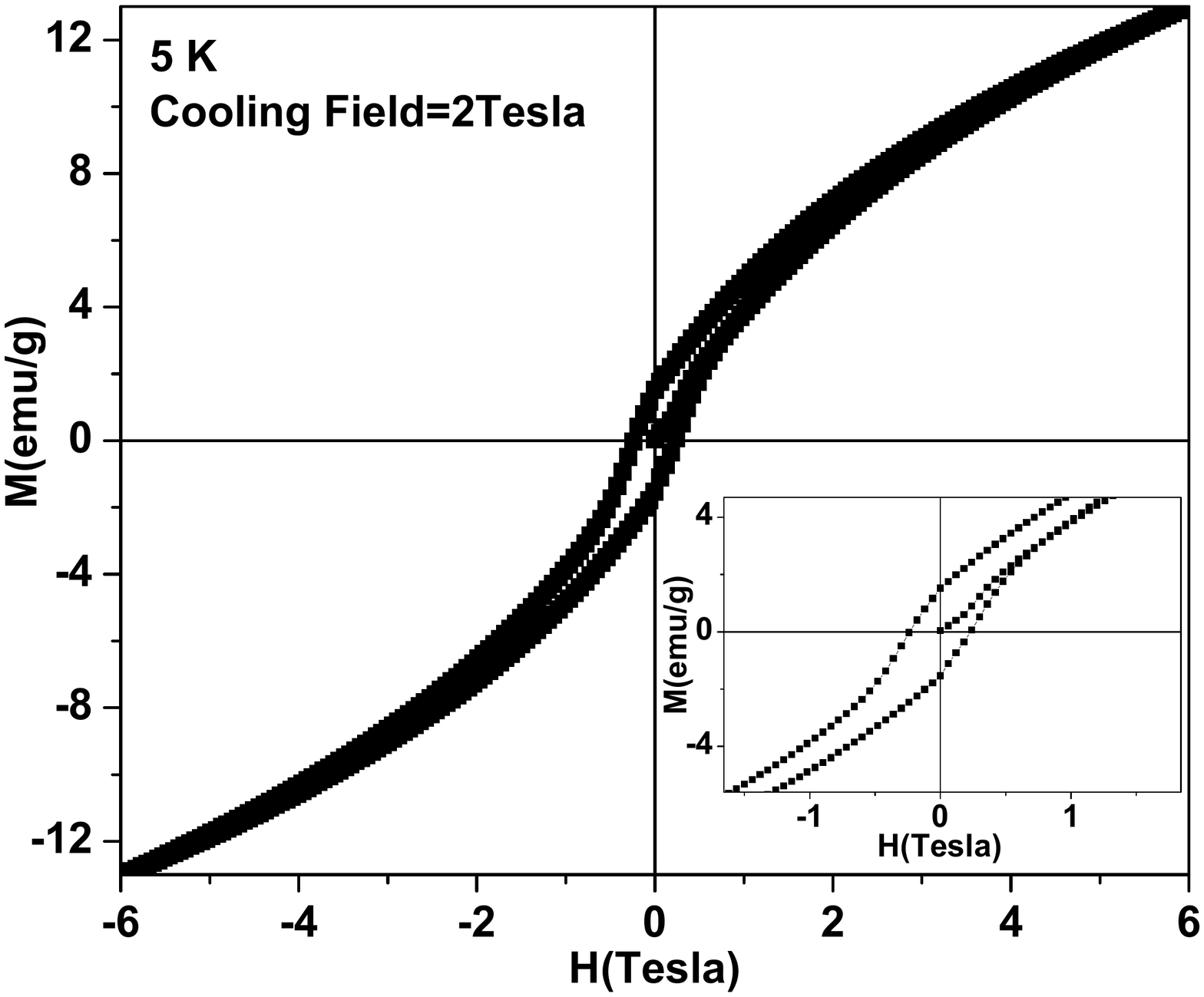}
\caption{Magnetization (M) vs. applied field (H) for BSMCO sample at 5 K cooled in the presence of 2Tesla field; inset of the figure shows an expanded view showing the absence of exchange bias.}
\label{Hys}
\end{figure}

Polycrystalline powder of $Bi_{0.5}Sr_{0.5}Mn_{0.9}Cr_{0.1}O_3$ (BSMCO) bulk sample was prepared via the solid state route. Calculated stoichiometric amounts of 99.9\% pure $Bi_2O_3$, $SrCO_3$, $MnCO_3$ and $Cr_2O_3$ were taken and mixed in an agate mortar and ground vigorously for an hour. This powder was then kept in a furnace for sintering at $900^0$ C for 12 hours (hrs). Sintered powder was taken out and ground again for an hour, pelleted and kept in the furnace twice, first at $1100^0$ C and then at $1300^0$ C (with intermediate grinding and pelleting) each for a duration of 24 hrs. The phase purity and the crystal structure of the sample was determined using X-ray diffractometer. Electron probe micro analysis (EPMA) was done to confirm the stoichiometric ratios. Magnetization measurements were done
using PPMS facility. EPR measurements were performed using Bruker EMX spectrometer at a nominal frequency of 9.4 GHz.

Room temperature X-ray diffraction pattern for BSMCO is shown in Figure~\ref{XRD}. Rietveld refinement was done on this XRD pattern taking reference data~\cite{Thakur} using the GSAS computer program. The sample crystallizes in tetragonal structure with space group P4mm, with cell parameters $a=b=3.9067 \AA$, $c=3.8010 \AA$ $\alpha=\beta=\gamma=90^0$,  with $\chi^2 = 8.563, Rp = 0.0219, wRp = 0.0509$. Stoichiometric ratios calculated using EPMA match satisfactorily with the expected values. The calculated stoichiometric ratios for elements are Bi=0.510, Sr=0.489, Mn=0.894, Cr=0.106. 

\section{MAGNETIZATION STUDIES}
The field cooled (FC) and the zero field cooled (ZFC) magnetization curves for BSMCO, measured under 100 Oe field, from 400 K till 4 K are shown in Figure~\ref{FZ}. Both FC and ZFC curves match with each other from 400 K till 50 K indicating the presence of PM phase; at $\sim$ 50 K there is a sharp increase in the FC magnetization curve indicating a building up of FM ordering and at the same temperature ZFC curve undergoes a peak. The large gap between the FC and ZFC curve points to short range FM ordering in the system~\cite{Itho}. We further note that the observed behavior of ZFC and FC magnetizations is very similar to that observed in $Bi_{1-x}Ca_xMnO_3$ (x=0.125) by  Woo et al.~\cite {Woo} and in $Ca_{0.9}Sm_{0.1}MnO_3$ by  Maignan et al.~\cite{Maignan}. Both these materials are argued to be exhibiting cluster glass phases.The inset to Figure~\ref{FZ} shows a plot of dM/dT vs. T on FC curve. Here a sharp dip at 50 K is due to the FM transition.

Figure~\ref{FZ} also presents the plot of $1/\chi_{dc}$ vs T fitted to the Curie-Weiss law  $\chi=C/(T+\theta)$. We get a value of C = 0.0148. Using this value of C we get $\mu_{eff}$ = 0.345 $\mu_B$ from $C = N_A(\mu_{eff})^2/3 k_B$. This is just $\sim$ 8\% of the ideal value of 4.282 $\mu_B$ expected if all the spins are ferromagnetically aligned. From this we infer that there is no long range FM in the sample and the sample consists of short range FM clusters. While the fit is seen to be good, the intercept on the x-axis is at +25 K, instead of a negative value expected if long range AF order is present. This result is understood, following Woo et al.~\cite{Woo}, who present an argument to explain the wrong sign in terms of the coexistence of AF and FM phases. According to their model the sign could be either positive or negative depending on the relative fractions of FM and AFM phases.

\begin{figure}[hbt!]
\includegraphics[width=\linewidth]{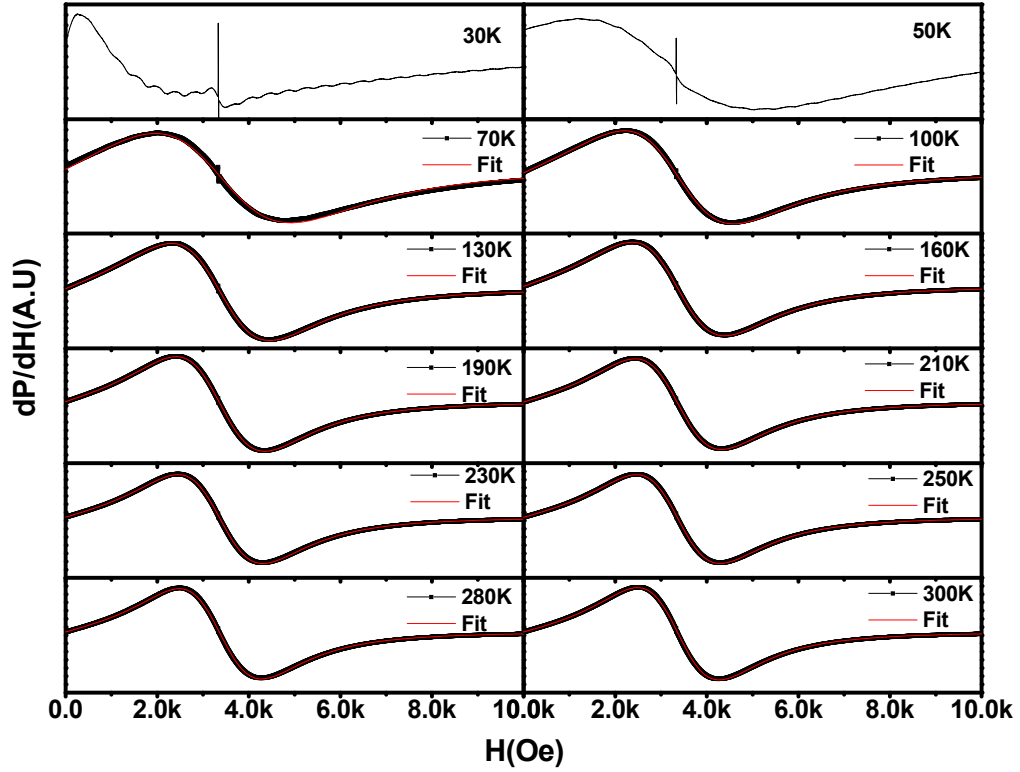}
\caption{ EPR spectra of BSMCO at different temperatures }
\label{EPR}
\end{figure}

\begin{figure}[hbt!]
\includegraphics[width=\linewidth]{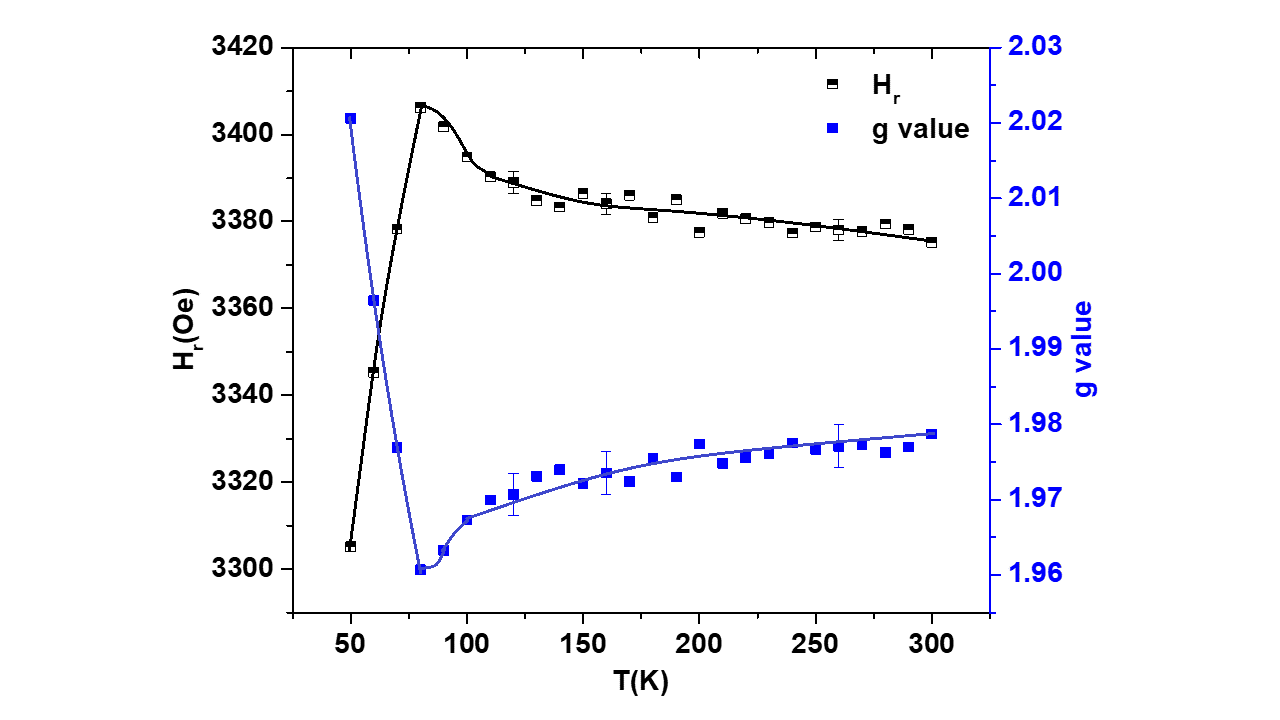}
\caption{ Temperature dependence of resonance field $H_r$ and g-value plot for the BSMCO sample (error bars are plotted at selected temperatures for both $H_r$ and g-value plots) }
\label{gH}
\end{figure}

We performed FC magnetization (M vs. H (applied field)) measurement at 5 K in the cooling field of 2 T. The plot is shown in Figure~\ref{Hys}. Closer inspection of the hysteresis plot (see the inset of Figure~\ref{Hys}) clearly indicates the presence of FM phase. Also note that the magnetization curve does not saturate even till 5Tesla field indicating the presence of non-ferromagnetic regions in the system, which may be due to either the presence of antiferromagnetically aligned spins or due to paramagnetic regions. The EPR results to be presented in the next section show that these regions are indeed antiferromagnetic.

\section{EPR STUDIES}
EPR spectra were recorded using a commercial X-band EPR spectrometer in the temperature range of 4 K to 300 K. A speck of DPPH was used as a field marker. The signals were fitted to the broad Lorentzian lineshape function in a derivative form~\cite{Joshi1},

\begin{equation}
\frac{dP}{dH} = A \frac{d}{dH} (\frac{\Delta H}{4(H-H_r)^2+(\Delta H)^2} + \frac{\Delta H}{4(H+H_r)^2+(\Delta H)^2})
%(dP/dH)=A  d/dH(∆H/(4(H-Ho)^(2 )+〖∆H〗^2 )+∆H/(4(H+Ho)^(2 )+〖∆H〗^2 ))
\end{equation}

where P is the microwave power absorbed by the sample at resonance, $\Delta H$ is the linewidth, $A$ is a quantity proportional to the intensity of the signal and $H_r$ is the resonance field, to extract the lineshape parameters intensity, g-value and linewidth. 

\begin{figure}[hbt!]
\includegraphics[width=\linewidth]{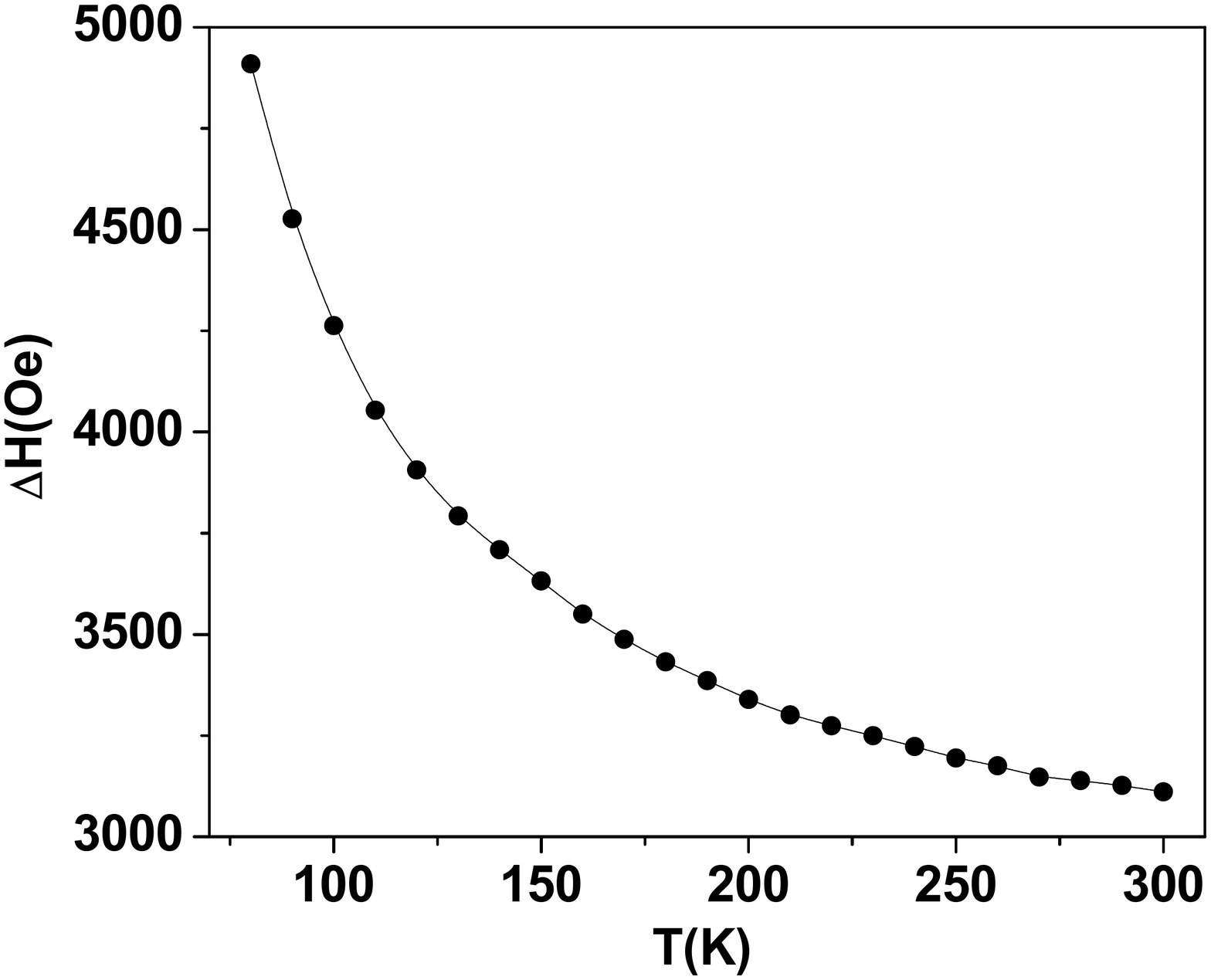}
\caption{ Temperature dependence of linewidth $\Delta H$ of the BSMCO sample }
\label{lw}
\end{figure}

\begin{figure}[hbt!]
\includegraphics[width=\linewidth]{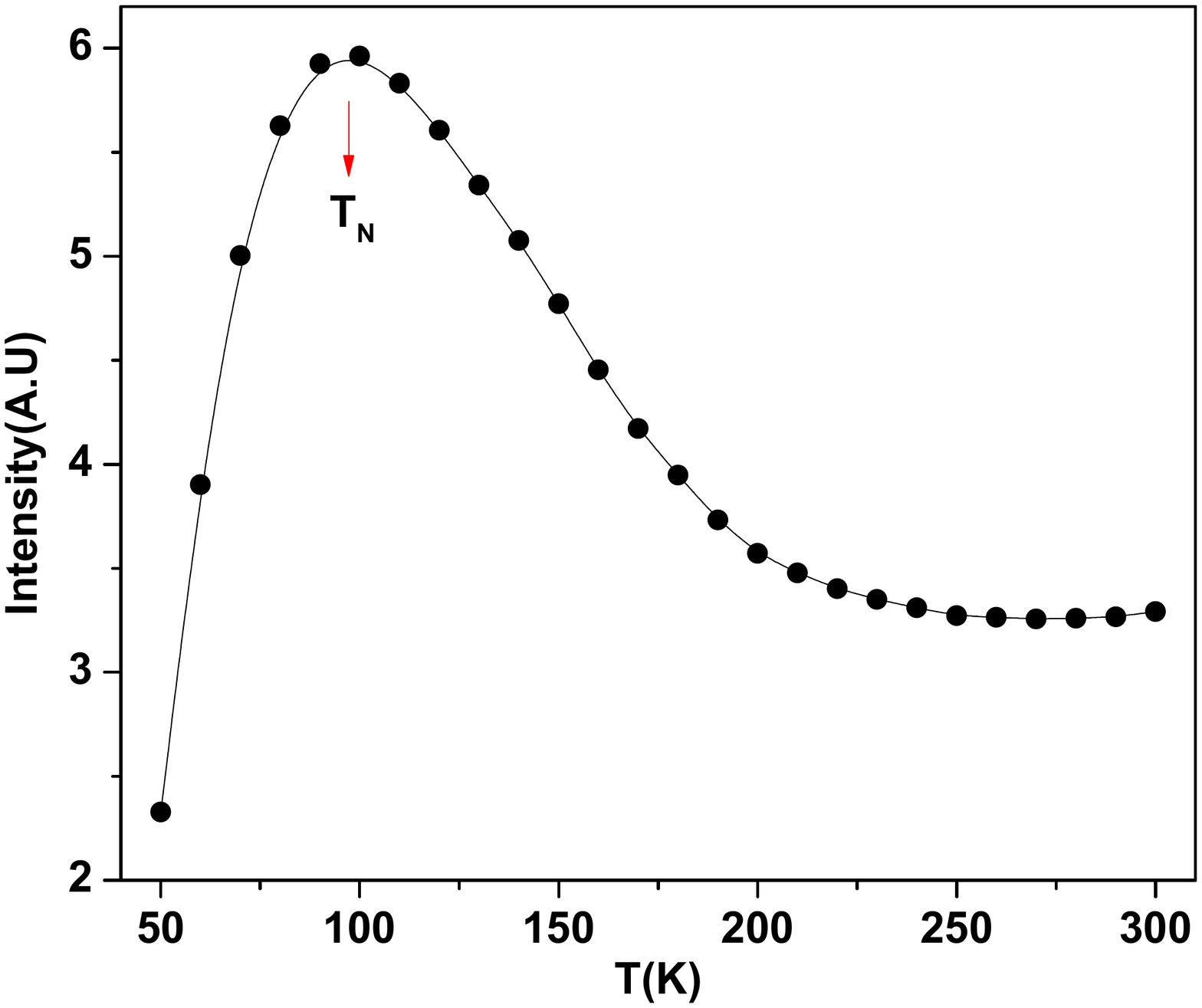}
\caption{ Temperature dependence of Intensity; arrow indicates the transition to an AFM phase at $\sim$ 110 K}
\label{Int}
\end{figure}

Figure~\ref{EPR} shows the EPR spectra at different temperatures for the BSMCO sample. Here the red line represents the double Lorentzian fit. Below 75 K we could not fit to the double Lorentzian lineshape because the signals were asymmetric owing to the presence of FM phase. But unlike bulk FM, where the signals become stronger here the signals get weaker and weaker with decreasing temperature. This suggests that the FM is not macroscopic but of short range in nature and also there is a competition between the ordered (FM) and disordered state; which conforms with our magnetization data. The lineshape parameters intensity I, linewidth $\Delta H_{pp} (= \Delta H/\sqrt 3)$ and resonance field ($H_r$) can be extracted from the fitted equation; but only for the symmetric signals. In case of asymmetric signals (given there is no loss of the signal below zero field), one can still measure the resonance field $H_r$ and peak to peak linewidth $\Delta H_{pp}$ and intensity (I) directly from the signal as described later. 

The temperature dependence of the resonance field $H_r$ and the g-value obtained from it are shown in Figure~\ref{gH}. Starting from room temperature, $H_r$ shows a slow increase with decreasing temperature till $\sim$75 K where it begins to decrease suddenly heralding the onset of ferromagnetic ordering and the development of local fields. The g-value, as expected shows a behavior opposite to this. 

%The g-value vs. temperature is plotted in the Figure~\ref{gH}. From the plot it is very clear that with decreasing temperature the g-value decreases till 75 K. At $\sim$ 75 K, we see a sudden increase in the g-value due to the building up of short range FM ordering. The resonance field $H_r$ vs. temperature for the sample is plotted in the inset of the Figure~\ref{gH}. The resonance field is constant in the paramagnetic region and decreases drastically as it approaches $T_c$. This can be attributed to the local field arising due to the transition.

The EPR linewidth behavior is as shown in Figure~\ref{lw}. As can be seen the linewidth increases with decreasing temperature going from PM phase to FM phase. This is unlike usual manganite systems where the linewidth decreases with decreasing temperature towards $T_c$, reaches a minimum at $T_{min} \sim$ 1.1 $T_c$ and begins to increase again as the temperature is decreased further~\cite{Sheng,Causa,Huber,Yuan}. To understand this atypical behavior better we tried fitting the linewidth variation to different known models~\cite{Causa,Huber,Seehra,Huber1,EAZ,Bhagat}. All these fittings resulted in the poor agreement with our data. The minimum in the linewidth observed at $\sim$ 1.1 $T_c$ in many manganites is a consequence of competing effects of spin-phonon interaction and FM/AFM fluctuations which lead to opposite dependence of the linewidth on the temperatures. An absence of an increase in the linewidth with increasing temperature indicates negligible contribution of spin-phonon interaction to the linewidth in our sample.

%We then calculated intensity using the formula~\cite{Seehra1},

%\begin{equation}
%I = (\Delta H_{pp})^2 h
%\end{equation}

%where $\Delta H_{pp}$ is the peak to peak linewidth and ‘h’ is peak to peak height.
%We measured ‘$\Delta H_{pp}$’ and ‘h’ directly from the signal and calculated the intensity down to 50 K. 
We calculated the intensity by double integrating the EPR signal. First we integrated the derivative curve to get the absorption signal; then we integrated the region from center of the absorption curve to the highest field in order to find the area under the curve. This value is multiplied twice to get the intensity of the EPR signal at different temperatures.
Figure~\ref{Int} shows the intensity plot against temperature. The EPR intensity is usually proportional to the DC susceptibility~\cite{Abragam}. Intensity vs. temperature is plotted and it is observed that intensity increases with decreasing temperature, closely resembling the FC magnetization trend as can be seen in the main Figure~\ref{FZ}. Intensity plot distinctly shows that as the temperature decreases intensity increases gradually peaking at around $\sim$ 110 K. This peak is very close to the AFM ordering temperature in an undoped system which is at $\sim$ 110 K~\cite{Herv,Kirste}. Earlier reports have shown that a peak is observed in an intensity plot if there exists an AFM ordering in the system~\cite{Rao,Joshi}. %Normalized `$I\times T$ vs. T' is plotted in the inset of Figure~\ref{Int}. As pointed in the figure there is a prominent peak at $\sim$ 110 K, which confirms the existence of AFM phase. 

\section{DISCUSSION}
10\% ‘Cr’ ion substitution has resulted in considerable changes in the system. $Cr^{3+}$ is the main driving force behind the appearance of FM phase as well as weakening of AFM phase in our system. We observe that our results are in agreement with earlier reports on Cr doped rare-earth manganites~\cite{Sharma,Dho,Troy1,Troy,Takuro} as well as Bi manganites~\cite{Xiong,Yamada}. Below we discuss the results of earlier reports on Cr substituted Bi manganite systems in comparison with our system.

According to Yamada et.al~\cite{Yamada} in $Bi_{1-x}Sr_xMn_{1-y}Cr_yO_3$ where x=0.25, 0.3, 0.4 and $0\leq  y \leq 0.1$, for all the samples Curie temperature is positive suggesting ferromagnetic interaction between Mn spin moments. The magnetization in the AFM phase increases with Cr doping upto 10\% and $T_N$ disappears for y=0.1. Thus the degree of AFM state by Cr doping decreases with increasing $x$. Authors also observe that upon Cr substitution the AFM phase in the system transforms into PM state rather than FM state. According to Xiong et.a~\cite{Xiong} in $Bi_{0.5}Ca_{0.5}Mn_{1-y}Cr_yO_3$ where $0 \leq y \leq 0.15$ with increasing Cr content; here for y=0, $T_N \sim$ 130 K. With increasing Cr content, AFM ordering shifts slightly to lower temperature but it is unaffected till $y \leq 0.24$. Above this there is a depression in the AFM ordering. The reciprocal susceptibilities of all the samples show positive $\theta_{cw}$ (curie constant) value, indicating the presence of local FM correlation and weakening
of AFM ordering. Our results are in agreement with earlier reports as in our sample we observe the weakening of AFM phase, emergence of FM phase, and below certain temperature both FM and AFM coexist together. According to magnetization short range FM ordering takes place below 50 K. EPR signals become asymmetric below 80 K indicating the presence of local field arising due to the presence of FM ordering, which is reflected in g-value plot. These asymmetric signals becoming weaker with decreasing temperature and absence of the minimum at $\sim$ $1.1T_c$ in the linewidth plot confirms the very nature of FM being short range. Magnetization does not display the presence of AFM phase but peak in the EPR intensity at $\sim$ 110 K exhibits the presence of AFM in the system.

We infer that at higher temperatures the sample is in the PM phase and as the temperature decreases the sample undergoes a transition to an AFM phase at $\sim$ 110 K due to the interaction between Mn ion moments. Further down the temperature at $\sim$  50 K, FM clusters are formed nucleating around the Cr ion moments leading to the formation of mixed phase. This causes increase in the magnetization. In this work we observe that 10\% of Cr substitution has resulted in the emergence of short range FM phase and weakening of AFM phase.

\section{CONCLUSION}
In summary, we prepared $Bi_{0.5}Sr_{0.5}Mn_{0.9}Cr_{0.1}O_3$ bulk sample and performed magnetization and EPR measurements. The FC and ZFC magnetization measurements show that the sample undergoes transition to a short range FM or cluster glass  phase at 50 K. The EPR data are consistent with the magnetization data by indicating the presence of short range FM. Further, EPR intensity gives the evidence of the presence of AFM below 110 K. We conclude that the sample below 50 K is in the mixed phase made up of FM clusters embedded in an AFM matrix.

\section*{Acknowledgment}
SVB thanks the National Academy of Sciences, India for the award of a Senior Scientist Platinum Jubilee Fellowship.

\bibliography{refer} 

\begin{thebibliography}{31}%
\makeatletter
\providecommand \@ifxundefined [1]{%
 \@ifx{#1\undefined}
}%
\providecommand \@ifnum [1]{%
 \ifnum #1\expandafter \@firstoftwo
 \else \expandafter \@secondoftwo
 \fi
}%
\providecommand \@ifx [1]{%
 \ifx #1\expandafter \@firstoftwo
 \else \expandafter \@secondoftwo
 \fi
}%
\providecommand \natexlab [1]{#1}%
\providecommand \enquote  [1]{``#1''}%
\providecommand \bibnamefont  [1]{#1}%
\providecommand \bibfnamefont [1]{#1}%
\providecommand \citenamefont [1]{#1}%
\providecommand \href@noop [0]{\@secondoftwo}%
\providecommand \href [0]{\begingroup \@sanitize@url \@href}%
\providecommand \@href[1]{\@@startlink{#1}\@@href}%
\providecommand \@@href[1]{\endgroup#1\@@endlink}%
\providecommand \@sanitize@url [0]{\catcode `\\12\catcode `\$12\catcode
  `\&12\catcode `\#12\catcode `\^12\catcode `\_12\catcode `\%12\relax}%
\providecommand \@@startlink[1]{}%
\providecommand \@@endlink[0]{}%
\providecommand \url  [0]{\begingroup\@sanitize@url \@url }%
\providecommand \@url [1]{\endgroup\@href {#1}{\urlprefix }}%
\providecommand \urlprefix  [0]{URL }%
\providecommand \Eprint [0]{\href }%
\providecommand \doibase [0]{http://dx.doi.org/}%
\providecommand \selectlanguage [0]{\@gobble}%
\providecommand \bibinfo  [0]{\@secondoftwo}%
\providecommand \bibfield  [0]{\@secondoftwo}%
\providecommand \translation [1]{[#1]}%
\providecommand \BibitemOpen [0]{}%
\providecommand \bibitemStop [0]{}%
\providecommand \bibitemNoStop [0]{.\EOS\space}%
\providecommand \EOS [0]{\spacefactor3000\relax}%
\providecommand \BibitemShut  [1]{\csname bibitem#1\endcsname}%
\let\auto@bib@innerbib\@empty
%</preamble>
\bibitem [{\citenamefont {Tokura}(2000)}]{tokura}%
  \BibitemOpen
  \bibfield  {author} {\bibinfo {author} {\bibfnamefont {Y.}~\bibnamefont
  {Tokura}},\ }\href@noop {} {\emph {\bibinfo {title} {Colossal
  magnetoresistive oxides}}}\ (\bibinfo  {publisher} {CRC Press},\ \bibinfo
  {year} {2000})\BibitemShut {NoStop}%
\bibitem [{\citenamefont {Rao}\ and\ \citenamefont {Raveau}(1998)}]{CNRrao}%
  \BibitemOpen
  \bibfield  {author} {\bibinfo {author} {\bibfnamefont {C.~N.~R.}\
  \bibnamefont {Rao}}\ and\ \bibinfo {author} {\bibfnamefont {B.}~\bibnamefont
  {Raveau}},\ }\href@noop {} {\emph {\bibinfo {title} {Colossal
  magnetoresistance, charge ordering and related properties of manganese
  oxides}}}\ (\bibinfo  {publisher} {World Scientific, Singapore},\ \bibinfo
  {year} {1998})\BibitemShut {NoStop}%
\bibitem [{\citenamefont {Ramakrishnan}(2007)}]{Rama}%
  \BibitemOpen
  \bibfield  {author} {\bibinfo {author} {\bibfnamefont {T.~V.}\ \bibnamefont
  {Ramakrishnan}},\ }\href {http://stacks.iop.org/0953-8984/19/i=12/a=125211}
  {\bibfield  {journal} {\bibinfo  {journal} {Journal of Physics: Condensed
  Matter}\ }\textbf {\bibinfo {volume} {19}},\ \bibinfo {pages} {125211}
  (\bibinfo {year} {2007})}\BibitemShut {NoStop}%
\bibitem [{\citenamefont {Dagotto}, \citenamefont {Hotta},\ and\ \citenamefont
  {Moreo}(2001)}]{Dagotto}%
  \BibitemOpen
  \bibfield  {author} {\bibinfo {author} {\bibfnamefont {E.}~\bibnamefont
  {Dagotto}}, \bibinfo {author} {\bibfnamefont {T.}~\bibnamefont {Hotta}}, \
  and\ \bibinfo {author} {\bibfnamefont {A.}~\bibnamefont {Moreo}},\ }\href
  {\doibase https://doi.org/10.1016/S0370-1573(00)00121-6} {\bibfield
  {journal} {\bibinfo  {journal} {Physics Reports}\ }\textbf {\bibinfo {volume}
  {344}},\ \bibinfo {pages} {1 } (\bibinfo {year} {2001})}\BibitemShut
  {NoStop}%
\bibitem [{\citenamefont {Rao}\ \emph {et~al.}(2005)\citenamefont {Rao},
  \citenamefont {Anuradha}, \citenamefont {Sarangi},\ and\ \citenamefont
  {Bhat}}]{Anu}%
  \BibitemOpen
  \bibfield  {author} {\bibinfo {author} {\bibfnamefont {S.~S.}\ \bibnamefont
  {Rao}}, \bibinfo {author} {\bibfnamefont {K.~N.}\ \bibnamefont {Anuradha}},
  \bibinfo {author} {\bibfnamefont {S.}~\bibnamefont {Sarangi}}, \ and\
  \bibinfo {author} {\bibfnamefont {S.~V.}\ \bibnamefont {Bhat}},\ }\href
  {\doibase 10.1063/1.2125129} {\bibfield  {journal} {\bibinfo  {journal}
  {Applied Physics Letters}\ }\textbf {\bibinfo {volume} {87}},\ \bibinfo
  {pages} {182503} (\bibinfo {year} {2005})}\BibitemShut {NoStop}%
\bibitem [{\citenamefont {Rao}\ \emph {et~al.}(2006)\citenamefont {Rao},
  \citenamefont {Tripathi}, \citenamefont {Pandey},\ and\ \citenamefont
  {Bhat}}]{Tripathi}%
  \BibitemOpen
  \bibfield  {author} {\bibinfo {author} {\bibfnamefont {S.~S.}\ \bibnamefont
  {Rao}}, \bibinfo {author} {\bibfnamefont {S.}~\bibnamefont {Tripathi}},
  \bibinfo {author} {\bibfnamefont {D.}~\bibnamefont {Pandey}}, \ and\ \bibinfo
  {author} {\bibfnamefont {S.~V.}\ \bibnamefont {Bhat}},\ }\href {\doibase
  10.1103/PhysRevB.74.144416} {\bibfield  {journal} {\bibinfo  {journal} {Phys.
  Rev. B}\ }\textbf {\bibinfo {volume} {74}},\ \bibinfo {pages} {144416}
  (\bibinfo {year} {2006})}\BibitemShut {NoStop}%
\bibitem [{\citenamefont {Sharma}\ and\ \citenamefont {Bhat}(2008)}]{Sharma}%
  \BibitemOpen
  \bibfield  {author} {\bibinfo {author} {\bibfnamefont {A.}~\bibnamefont
  {Sharma}}\ and\ \bibinfo {author} {\bibfnamefont {S.~V.}\ \bibnamefont
  {Bhat}},\ }\href {\doibase 10.1007/s00723-008-0050-7} {\bibfield  {journal}
  {\bibinfo  {journal} {Applied Magnetic Resonance}\ }\textbf {\bibinfo
  {volume} {33}},\ \bibinfo {pages} {11} (\bibinfo {year} {2008})}\BibitemShut
  {NoStop}%
\bibitem [{\citenamefont {Xiong}\ \emph {et~al.}(2004)\citenamefont {Xiong},
  \citenamefont {Sun}, \citenamefont {Li}, \citenamefont {Zhang}, \citenamefont
  {Zhao},\ and\ \citenamefont {Shen}}]{Xiong}%
  \BibitemOpen
  \bibfield  {author} {\bibinfo {author} {\bibfnamefont {C.~M.}\ \bibnamefont
  {Xiong}}, \bibinfo {author} {\bibfnamefont {J.~R.}\ \bibnamefont {Sun}},
  \bibinfo {author} {\bibfnamefont {R.~W.}\ \bibnamefont {Li}}, \bibinfo
  {author} {\bibfnamefont {S.~Y.}\ \bibnamefont {Zhang}}, \bibinfo {author}
  {\bibfnamefont {T.~Y.}\ \bibnamefont {Zhao}}, \ and\ \bibinfo {author}
  {\bibfnamefont {B.~G.}\ \bibnamefont {Shen}},\ }\href {\doibase
  http://dx.doi.org/10.1063/1.1636514} {\bibfield  {journal} {\bibinfo
  {journal} {Journal of Applied Physics}\ }\textbf {\bibinfo {volume} {95}},\
  \bibinfo {pages} {1336} (\bibinfo {year} {2004})}\BibitemShut {NoStop}%
\bibitem [{\citenamefont {Dho}, \citenamefont {Kim},\ and\ \citenamefont
  {Hur}(2002)}]{Dho}%
  \BibitemOpen
  \bibfield  {author} {\bibinfo {author} {\bibfnamefont {J.}~\bibnamefont
  {Dho}}, \bibinfo {author} {\bibfnamefont {W.~S.}\ \bibnamefont {Kim}}, \ and\
  \bibinfo {author} {\bibfnamefont {N.~H.}\ \bibnamefont {Hur}},\ }\href
  {\doibase 10.1103/PhysRevLett.89.027202} {\bibfield  {journal} {\bibinfo
  {journal} {Phys. Rev. Lett.}\ }\textbf {\bibinfo {volume} {89}},\ \bibinfo
  {pages} {027202} (\bibinfo {year} {2002})}\BibitemShut {NoStop}%
\bibitem [{\citenamefont {Troyanchuk}\ \emph {et~al.}(2002)\citenamefont
  {Troyanchuk}, \citenamefont {Bushinsky}, \citenamefont {Eremenko},
  \citenamefont {Sirenko},\ and\ \citenamefont {Szymczak}}]{Troy1}%
  \BibitemOpen
  \bibfield  {author} {\bibinfo {author} {\bibfnamefont {I.~O.}\ \bibnamefont
  {Troyanchuk}}, \bibinfo {author} {\bibfnamefont {M.~V.}\ \bibnamefont
  {Bushinsky}}, \bibinfo {author} {\bibfnamefont {V.~V.}\ \bibnamefont
  {Eremenko}}, \bibinfo {author} {\bibfnamefont {V.~A.}\ \bibnamefont
  {Sirenko}}, \ and\ \bibinfo {author} {\bibfnamefont {H.}~\bibnamefont
  {Szymczak}},\ }\href {\doibase http://dx.doi.org/10.1063/1.1449184}
  {\bibfield  {journal} {\bibinfo  {journal} {Low Temperature Physics}\
  }\textbf {\bibinfo {volume} {28}},\ \bibinfo {pages} {45} (\bibinfo {year}
  {2002})}\BibitemShut {NoStop}%
\bibitem [{\citenamefont {Troyanchuk}\ \emph {et~al.}(2006)\citenamefont
  {Troyanchuk}, \citenamefont {Pushkarev}, \citenamefont {Bushinskiĭ},\ and\
  \citenamefont {Hamari-Sile}}]{Troy}%
  \BibitemOpen
  \bibfield  {author} {\bibinfo {author} {\bibfnamefont {I.}~\bibnamefont
  {Troyanchuk}}, \bibinfo {author} {\bibfnamefont {N.}~\bibnamefont
  {Pushkarev}}, \bibinfo {author} {\bibfnamefont {M.}~\bibnamefont
  {Bushinskiĭ}}, \ and\ \bibinfo {author} {\bibfnamefont {E.}~\bibnamefont
  {Hamari-Sile}},\ }\href {\doibase 10.1134/S106378340607016X} {\bibfield
  {journal} {\bibinfo  {journal} {Physics of the Solid State}\ }\textbf
  {\bibinfo {volume} {48}},\ \bibinfo {pages} {1315} (\bibinfo {year}
  {2006})}\BibitemShut {NoStop}%
\bibitem [{\citenamefont {Katsufuji}\ \emph {et~al.}(1999)\citenamefont
  {Katsufuji}, \citenamefont {Cheong}, \citenamefont {Mori},\ and\
  \citenamefont {Chen}}]{Takuro}%
  \BibitemOpen
  \bibfield  {author} {\bibinfo {author} {\bibfnamefont {T.}~\bibnamefont
  {Katsufuji}}, \bibinfo {author} {\bibfnamefont {S.-W.}\ \bibnamefont
  {Cheong}}, \bibinfo {author} {\bibfnamefont {S.}~\bibnamefont {Mori}}, \ and\
  \bibinfo {author} {\bibfnamefont {C.-H.}\ \bibnamefont {Chen}},\ }\href@noop
  {} {\bibfield  {journal} {\bibinfo  {journal} {Journal of the Physical
  Society of Japan}\ }\textbf {\bibinfo {volume} {68}},\ \bibinfo {pages}
  {1090} (\bibinfo {year} {1999})}\BibitemShut {NoStop}%
\bibitem [{\citenamefont {Yamada}, \citenamefont {Sugano},\ and\ \citenamefont
  {hisa Arima}(2006)}]{Yamada}%
  \BibitemOpen
  \bibfield  {author} {\bibinfo {author} {\bibfnamefont {S.}~\bibnamefont
  {Yamada}}, \bibinfo {author} {\bibfnamefont {E.}~\bibnamefont {Sugano}}, \
  and\ \bibinfo {author} {\bibfnamefont {T.}~\bibnamefont {hisa Arima}},\
  }\href {\doibase http://dx.doi.org/10.1016/j.jssc.2006.05.045} {\bibfield
  {journal} {\bibinfo  {journal} {Journal of Solid State Chemistry}\ }\textbf
  {\bibinfo {volume} {179}},\ \bibinfo {pages} {3121 } (\bibinfo {year}
  {2006})}\BibitemShut {NoStop}%
\bibitem [{\citenamefont {Hervieu}\ \emph {et~al.}(2001)\citenamefont
  {Hervieu}, \citenamefont {Maignan}, \citenamefont {Martin}, \citenamefont
  {Nguyen},\ and\ \citenamefont {Raveau}}]{Herv}%
  \BibitemOpen
  \bibfield  {author} {\bibinfo {author} {\bibfnamefont {M.}~\bibnamefont
  {Hervieu}}, \bibinfo {author} {\bibfnamefont {A.}~\bibnamefont {Maignan}},
  \bibinfo {author} {\bibfnamefont {C.}~\bibnamefont {Martin}}, \bibinfo
  {author} {\bibfnamefont {N.}~\bibnamefont {Nguyen}}, \ and\ \bibinfo {author}
  {\bibfnamefont {B.}~\bibnamefont {Raveau}},\ }\href@noop {} {\bibfield
  {journal} {\bibinfo  {journal} {Chemistry of Materials}\ }\textbf {\bibinfo
  {volume} {13}},\ \bibinfo {pages} {1356} (\bibinfo {year}
  {2001})}\BibitemShut {NoStop}%
\bibitem [{\citenamefont {Kirste}\ \emph {et~al.}(2003)\citenamefont {Kirste},
  \citenamefont {Goiran}, \citenamefont {Respaud}, \citenamefont {Vanaken},
  \citenamefont {Broto}, \citenamefont {Rakoto}, \citenamefont {von Ortenberg},
  \citenamefont {Frontera},\ and\ \citenamefont
  {Garc\'{\i}a-Mu\~noz}}]{Kirste}%
  \BibitemOpen
  \bibfield  {author} {\bibinfo {author} {\bibfnamefont {A.}~\bibnamefont
  {Kirste}}, \bibinfo {author} {\bibfnamefont {M.}~\bibnamefont {Goiran}},
  \bibinfo {author} {\bibfnamefont {M.}~\bibnamefont {Respaud}}, \bibinfo
  {author} {\bibfnamefont {J.}~\bibnamefont {Vanaken}}, \bibinfo {author}
  {\bibfnamefont {J.~M.}\ \bibnamefont {Broto}}, \bibinfo {author}
  {\bibfnamefont {H.}~\bibnamefont {Rakoto}}, \bibinfo {author} {\bibfnamefont
  {M.}~\bibnamefont {von Ortenberg}}, \bibinfo {author} {\bibfnamefont
  {C.}~\bibnamefont {Frontera}}, \ and\ \bibinfo {author} {\bibfnamefont
  {J.~L.}\ \bibnamefont {Garc\'{\i}a-Mu\~noz}},\ }\href {\doibase
  10.1103/PhysRevB.67.134413} {\bibfield  {journal} {\bibinfo  {journal} {Phys.
  Rev. B}\ }\textbf {\bibinfo {volume} {67}},\ \bibinfo {pages} {134413}
  (\bibinfo {year} {2003})}\BibitemShut {NoStop}%
\bibitem [{\citenamefont {Rao}\ and\ \citenamefont {Bhat}(2007)}]{Rao}%
  \BibitemOpen
  \bibfield  {author} {\bibinfo {author} {\bibfnamefont {S.~S.}\ \bibnamefont
  {Rao}}\ and\ \bibinfo {author} {\bibfnamefont {S.~V.}\ \bibnamefont {Bhat}},\
  }\href {\doibase doi:10.1166/jnn.2007.762} {\bibfield  {journal} {\bibinfo
  {journal} {Journal of Nanoscience and Nanotechnology}\ }\textbf {\bibinfo
  {volume} {7}},\ \bibinfo {pages} {2025} (\bibinfo {year} {2007})}\BibitemShut
  {NoStop}%
\bibitem [{\citenamefont {Thakur}, \citenamefont {Pandey},\ and\ \citenamefont
  {Singh}(2013)}]{Thakur}%
  \BibitemOpen
  \bibfield  {author} {\bibinfo {author} {\bibfnamefont {S.}~\bibnamefont
  {Thakur}}, \bibinfo {author} {\bibfnamefont {O.}~\bibnamefont {Pandey}}, \
  and\ \bibinfo {author} {\bibfnamefont {K.}~\bibnamefont {Singh}},\ }\href
  {\doibase http://dx.doi.org/10.1016/j.ceramint.2013.01.035} {\bibfield
  {journal} {\bibinfo  {journal} {Ceramics International}\ }\textbf {\bibinfo
  {volume} {39}},\ \bibinfo {pages} {6165 } (\bibinfo {year}
  {2013})}\BibitemShut {NoStop}%
\bibitem [{\citenamefont {Itoh}\ \emph {et~al.}(1994)\citenamefont {Itoh},
  \citenamefont {Natori}, \citenamefont {Kubota},\ and\ \citenamefont
  {Motoya}}]{Itho}%
  \BibitemOpen
  \bibfield  {author} {\bibinfo {author} {\bibfnamefont {M.}~\bibnamefont
  {Itoh}}, \bibinfo {author} {\bibfnamefont {I.}~\bibnamefont {Natori}},
  \bibinfo {author} {\bibfnamefont {S.}~\bibnamefont {Kubota}}, \ and\ \bibinfo
  {author} {\bibfnamefont {K.}~\bibnamefont {Motoya}},\ }\href@noop {}
  {\bibfield  {journal} {\bibinfo  {journal} {Journal of the Physical Society
  of Japan}\ }\textbf {\bibinfo {volume} {63}},\ \bibinfo {pages} {1486}
  (\bibinfo {year} {1994})}\BibitemShut {NoStop}%
\bibitem [{\citenamefont {Woo}\ \emph {et~al.}(2004)\citenamefont {Woo},
  \citenamefont {Tyson}, \citenamefont {Croft},\ and\ \citenamefont
  {Cheong}}]{Woo}%
  \BibitemOpen
  \bibfield  {author} {\bibinfo {author} {\bibfnamefont {H.}~\bibnamefont
  {Woo}}, \bibinfo {author} {\bibfnamefont {T.~A.}\ \bibnamefont {Tyson}},
  \bibinfo {author} {\bibfnamefont {M.}~\bibnamefont {Croft}}, \ and\ \bibinfo
  {author} {\bibfnamefont {S.-W.}\ \bibnamefont {Cheong}},\ }\href {\doibase
  10.1088/0953-8984/16/15/020} {\bibfield  {journal} {\bibinfo  {journal}
  {Journal of Physics: Condensed Matter}\ }\textbf {\bibinfo {volume} {16}},\
  \bibinfo {pages} {2689} (\bibinfo {year} {2004})}\BibitemShut {NoStop}%
\bibitem [{\citenamefont {Maignan}\ \emph {et~al.}(1998)\citenamefont
  {Maignan}, \citenamefont {Martin}, \citenamefont {Damay}, \citenamefont
  {Raveau},\ and\ \citenamefont {Hejtmanek}}]{Maignan}%
  \BibitemOpen
  \bibfield  {author} {\bibinfo {author} {\bibfnamefont {A.}~\bibnamefont
  {Maignan}}, \bibinfo {author} {\bibfnamefont {C.}~\bibnamefont {Martin}},
  \bibinfo {author} {\bibfnamefont {F.}~\bibnamefont {Damay}}, \bibinfo
  {author} {\bibfnamefont {B.}~\bibnamefont {Raveau}}, \ and\ \bibinfo {author}
  {\bibfnamefont {J.}~\bibnamefont {Hejtmanek}},\ }\href {\doibase
  10.1103/PhysRevB.58.2758} {\bibfield  {journal} {\bibinfo  {journal} {Phys.
  Rev. B}\ }\textbf {\bibinfo {volume} {58}},\ \bibinfo {pages} {2758}
  (\bibinfo {year} {1998})}\BibitemShut {NoStop}%
\bibitem [{\citenamefont {Joshi}\ and\ \citenamefont {Bhat}(2004)}]{Joshi1}%
  \BibitemOpen
  \bibfield  {author} {\bibinfo {author} {\bibfnamefont {J.~P.}\ \bibnamefont
  {Joshi}}\ and\ \bibinfo {author} {\bibfnamefont {S.~V.}\ \bibnamefont
  {Bhat}},\ }\href {\doibase http://dx.doi.org/10.1016/j.jmr.2004.03.018}
  {\bibfield  {journal} {\bibinfo  {journal} {Journal of Magnetic Resonance}\
  }\textbf {\bibinfo {volume} {168}},\ \bibinfo {pages} {284 } (\bibinfo {year}
  {2004})}\BibitemShut {NoStop}%
\bibitem [{\citenamefont {Shengelaya}\ \emph {et~al.}(1996)\citenamefont
  {Shengelaya}, \citenamefont {Zhao}, \citenamefont {Keller},\ and\
  \citenamefont {M\"uller}}]{Sheng}%
  \BibitemOpen
  \bibfield  {author} {\bibinfo {author} {\bibfnamefont {A.}~\bibnamefont
  {Shengelaya}}, \bibinfo {author} {\bibfnamefont {G.-m.}\ \bibnamefont
  {Zhao}}, \bibinfo {author} {\bibfnamefont {H.}~\bibnamefont {Keller}}, \ and\
  \bibinfo {author} {\bibfnamefont {K.~A.}\ \bibnamefont {M\"uller}},\ }\href
  {\doibase 10.1103/PhysRevLett.77.5296} {\bibfield  {journal} {\bibinfo
  {journal} {Phys. Rev. Lett.}\ }\textbf {\bibinfo {volume} {77}},\ \bibinfo
  {pages} {5296} (\bibinfo {year} {1996})}\BibitemShut {NoStop}%
\bibitem [{\citenamefont {Causa}\ \emph {et~al.}(1998)\citenamefont {Causa},
  \citenamefont {Tovar}, \citenamefont {Caneiro}, \citenamefont {Prado},
  \citenamefont {Iba\~nez}, \citenamefont {Ramos}, \citenamefont {Butera},
  \citenamefont {Alascio}, \citenamefont {Obradors}, \citenamefont {Pi\~nol},
  \citenamefont {Rivadulla}, \citenamefont {V\'azquez-V\'azquez}, \citenamefont
  {L\'opez-Quintela}, \citenamefont {Rivas}, \citenamefont {Tokura},\ and\
  \citenamefont {Oseroff}}]{Causa}%
  \BibitemOpen
  \bibfield  {author} {\bibinfo {author} {\bibfnamefont {M.~T.}\ \bibnamefont
  {Causa}}, \bibinfo {author} {\bibfnamefont {M.}~\bibnamefont {Tovar}},
  \bibinfo {author} {\bibfnamefont {A.}~\bibnamefont {Caneiro}}, \bibinfo
  {author} {\bibfnamefont {F.}~\bibnamefont {Prado}}, \bibinfo {author}
  {\bibfnamefont {G.}~\bibnamefont {Iba\~nez}}, \bibinfo {author}
  {\bibfnamefont {C.~A.}\ \bibnamefont {Ramos}}, \bibinfo {author}
  {\bibfnamefont {A.}~\bibnamefont {Butera}}, \bibinfo {author} {\bibfnamefont
  {B.}~\bibnamefont {Alascio}}, \bibinfo {author} {\bibfnamefont
  {X.}~\bibnamefont {Obradors}}, \bibinfo {author} {\bibfnamefont
  {S.}~\bibnamefont {Pi\~nol}}, \bibinfo {author} {\bibfnamefont
  {F.}~\bibnamefont {Rivadulla}}, \bibinfo {author} {\bibfnamefont
  {C.}~\bibnamefont {V\'azquez-V\'azquez}}, \bibinfo {author} {\bibfnamefont
  {M.~A.}\ \bibnamefont {L\'opez-Quintela}}, \bibinfo {author} {\bibfnamefont
  {J.}~\bibnamefont {Rivas}}, \bibinfo {author} {\bibfnamefont
  {Y.}~\bibnamefont {Tokura}}, \ and\ \bibinfo {author} {\bibfnamefont {S.~B.}\
  \bibnamefont {Oseroff}},\ }\href {\doibase 10.1103/PhysRevB.58.3233}
  {\bibfield  {journal} {\bibinfo  {journal} {Phys. Rev. B}\ }\textbf {\bibinfo
  {volume} {58}},\ \bibinfo {pages} {3233} (\bibinfo {year}
  {1998})}\BibitemShut {NoStop}%
\bibitem [{\citenamefont {Huber}\ \emph {et~al.}(1999)\citenamefont {Huber},
  \citenamefont {Alejandro}, \citenamefont {Caneiro}, \citenamefont {Causa},
  \citenamefont {Prado}, \citenamefont {Tovar},\ and\ \citenamefont
  {Oseroff}}]{Huber}%
  \BibitemOpen
  \bibfield  {author} {\bibinfo {author} {\bibfnamefont {D.~L.}\ \bibnamefont
  {Huber}}, \bibinfo {author} {\bibfnamefont {G.}~\bibnamefont {Alejandro}},
  \bibinfo {author} {\bibfnamefont {A.}~\bibnamefont {Caneiro}}, \bibinfo
  {author} {\bibfnamefont {M.~T.}\ \bibnamefont {Causa}}, \bibinfo {author}
  {\bibfnamefont {F.}~\bibnamefont {Prado}}, \bibinfo {author} {\bibfnamefont
  {M.}~\bibnamefont {Tovar}}, \ and\ \bibinfo {author} {\bibfnamefont {S.~B.}\
  \bibnamefont {Oseroff}},\ }\href {\doibase 10.1103/PhysRevB.60.12155}
  {\bibfield  {journal} {\bibinfo  {journal} {Phys. Rev. B}\ }\textbf {\bibinfo
  {volume} {60}},\ \bibinfo {pages} {12155} (\bibinfo {year}
  {1999})}\BibitemShut {NoStop}%
\bibitem [{\citenamefont {Yuan}\ \emph {et~al.}(2000)\citenamefont {Yuan},
  \citenamefont {Li}, \citenamefont {Jiang}, \citenamefont {Li}, \citenamefont
  {Zeng}, \citenamefont {Yang}, \citenamefont {Huang},\ and\ \citenamefont
  {Jin}}]{Yuan}%
  \BibitemOpen
  \bibfield  {author} {\bibinfo {author} {\bibfnamefont {S.~L.}\ \bibnamefont
  {Yuan}}, \bibinfo {author} {\bibfnamefont {G.}~\bibnamefont {Li}}, \bibinfo
  {author} {\bibfnamefont {Y.}~\bibnamefont {Jiang}}, \bibinfo {author}
  {\bibfnamefont {J.~Q.}\ \bibnamefont {Li}}, \bibinfo {author} {\bibfnamefont
  {X.~Y.}\ \bibnamefont {Zeng}}, \bibinfo {author} {\bibfnamefont {Y.~P.}\
  \bibnamefont {Yang}}, \bibinfo {author} {\bibfnamefont {Z.}~\bibnamefont
  {Huang}}, \ and\ \bibinfo {author} {\bibfnamefont {S.~Z.}\ \bibnamefont
  {Jin}},\ }\href {http://stacks.iop.org/0953-8984/12/i=6/a=107} {\bibfield
  {journal} {\bibinfo  {journal} {Journal of Physics: Condensed Matter}\
  }\textbf {\bibinfo {volume} {12}},\ \bibinfo {pages} {L109} (\bibinfo {year}
  {2000})}\BibitemShut {NoStop}%
\bibitem [{\citenamefont {Seehra}\ and\ \citenamefont {Jr.}(1970)}]{Seehra}%
  \BibitemOpen
  \bibfield  {author} {\bibinfo {author} {\bibfnamefont {M.~S.}\ \bibnamefont
  {Seehra}}\ and\ \bibinfo {author} {\bibfnamefont {T.~C.}\ \bibnamefont
  {Jr.}},\ }\href {\doibase http://dx.doi.org/10.1016/0038-1098(70)90432-1}
  {\bibfield  {journal} {\bibinfo  {journal} {Solid State Communications}\
  }\textbf {\bibinfo {volume} {8}},\ \bibinfo {pages} {787 } (\bibinfo {year}
  {1970})}\BibitemShut {NoStop}%
\bibitem [{\citenamefont {Huber}(1972)}]{Huber1}%
  \BibitemOpen
  \bibfield  {author} {\bibinfo {author} {\bibfnamefont {D.~L.}\ \bibnamefont
  {Huber}},\ }\href {\doibase 10.1103/PhysRevB.6.3180} {\bibfield  {journal}
  {\bibinfo  {journal} {Phys. Rev. B}\ }\textbf {\bibinfo {volume} {6}},\
  \bibinfo {pages} {3180} (\bibinfo {year} {1972})}\BibitemShut {NoStop}%
\bibitem [{\citenamefont {Zvereva}\ \emph {et~al.}(2015)\citenamefont
  {Zvereva}, \citenamefont {Stratan}, \citenamefont {Ovchenkov}, \citenamefont
  {Nalbandyan}, \citenamefont {Lin}, \citenamefont {Vavilova}, \citenamefont
  {Iakovleva}, \citenamefont {Abdel-Hafiez}, \citenamefont {Silhanek},
  \citenamefont {Chen}, \citenamefont {Stroppa}, \citenamefont {Picozzi},
  \citenamefont {Jeschke}, \citenamefont {Valent\'{\i}},\ and\ \citenamefont
  {Vasiliev}}]{EAZ}%
  \BibitemOpen
  \bibfield  {author} {\bibinfo {author} {\bibfnamefont {E.~A.}\ \bibnamefont
  {Zvereva}}, \bibinfo {author} {\bibfnamefont {M.~I.}\ \bibnamefont
  {Stratan}}, \bibinfo {author} {\bibfnamefont {Y.~A.}\ \bibnamefont
  {Ovchenkov}}, \bibinfo {author} {\bibfnamefont {V.~B.}\ \bibnamefont
  {Nalbandyan}}, \bibinfo {author} {\bibfnamefont {J.-Y.}\ \bibnamefont {Lin}},
  \bibinfo {author} {\bibfnamefont {E.~L.}\ \bibnamefont {Vavilova}}, \bibinfo
  {author} {\bibfnamefont {M.~F.}\ \bibnamefont {Iakovleva}}, \bibinfo {author}
  {\bibfnamefont {M.}~\bibnamefont {Abdel-Hafiez}}, \bibinfo {author}
  {\bibfnamefont {A.~V.}\ \bibnamefont {Silhanek}}, \bibinfo {author}
  {\bibfnamefont {X.-J.}\ \bibnamefont {Chen}}, \bibinfo {author}
  {\bibfnamefont {A.}~\bibnamefont {Stroppa}}, \bibinfo {author} {\bibfnamefont
  {S.}~\bibnamefont {Picozzi}}, \bibinfo {author} {\bibfnamefont {H.~O.}\
  \bibnamefont {Jeschke}}, \bibinfo {author} {\bibfnamefont {R.}~\bibnamefont
  {Valent\'{\i}}}, \ and\ \bibinfo {author} {\bibfnamefont {A.~N.}\
  \bibnamefont {Vasiliev}},\ }\href {\doibase 10.1103/PhysRevB.92.144401}
  {\bibfield  {journal} {\bibinfo  {journal} {Phys. Rev. B}\ }\textbf {\bibinfo
  {volume} {92}},\ \bibinfo {pages} {144401} (\bibinfo {year}
  {2015})}\BibitemShut {NoStop}%
\bibitem [{\citenamefont {Bhagat}, \citenamefont {Spano},\ and\ \citenamefont
  {Lloyd}(1981)}]{Bhagat}%
  \BibitemOpen
  \bibfield  {author} {\bibinfo {author} {\bibfnamefont {S.}~\bibnamefont
  {Bhagat}}, \bibinfo {author} {\bibfnamefont {M.}~\bibnamefont {Spano}}, \
  and\ \bibinfo {author} {\bibfnamefont {J.}~\bibnamefont {Lloyd}},\ }\href
  {\doibase https://doi.org/10.1016/0038-1098(81)90457-9} {\bibfield  {journal}
  {\bibinfo  {journal} {Solid State Communications}\ }\textbf {\bibinfo
  {volume} {38}},\ \bibinfo {pages} {261 } (\bibinfo {year}
  {1981})}\BibitemShut {NoStop}%
\bibitem [{\citenamefont {Abragam}(2006)}]{Abragam}%
  \BibitemOpen
  \bibfield  {author} {\bibinfo {author} {\bibfnamefont {A.}~\bibnamefont
  {Abragam}},\ }\href@noop {} {\emph {\bibinfo {title} {Principles of nuclear
  magnetic resonance}}}\ (\bibinfo  {publisher} {Clarendon Press, Oxford},\
  \bibinfo {year} {2006})\ pp.\ \bibinfo {pages} {40--42}\BibitemShut {NoStop}%
\bibitem [{\citenamefont {Joshi}\ \emph {et~al.}(2001)\citenamefont {Joshi},
  \citenamefont {Gupta}, \citenamefont {Sood}, \citenamefont {Bhat},
  \citenamefont {Raju},\ and\ \citenamefont {Rao}}]{Joshi}%
  \BibitemOpen
  \bibfield  {author} {\bibinfo {author} {\bibfnamefont {J.~P.}\ \bibnamefont
  {Joshi}}, \bibinfo {author} {\bibfnamefont {R.}~\bibnamefont {Gupta}},
  \bibinfo {author} {\bibfnamefont {A.~K.}\ \bibnamefont {Sood}}, \bibinfo
  {author} {\bibfnamefont {S.~V.}\ \bibnamefont {Bhat}}, \bibinfo {author}
  {\bibfnamefont {A.~R.}\ \bibnamefont {Raju}}, \ and\ \bibinfo {author}
  {\bibfnamefont {C.~N.~R.}\ \bibnamefont {Rao}},\ }\href {\doibase
  10.1103/PhysRevB.65.024410} {\bibfield  {journal} {\bibinfo  {journal} {Phys.
  Rev. B}\ }\textbf {\bibinfo {volume} {65}},\ \bibinfo {pages} {024410}
  (\bibinfo {year} {2001})}\BibitemShut {NoStop}%
\end{thebibliography}%

\end{document}